# BCS quantum critical phenomena

Yong Tao[†]

College of Economics and Management, Southwest University, Chongqing, China

**Abstract:** Theoretically, we recently showed that the scaling relation between the transition temperature $T_c$ and the superfluid density at zero temperature $n_s(0)$ might exhibit a parabolic pattern: $T_c \propto \sqrt{n_s(0)}$ [Scientific Reports **6** (2016) 23863]. It is significantly different from the linear scaling described by Homes' law, which is well known as a mean-field result. More recently, Božović *et al.* have observed such a parabolic scaling in the overdoped copper oxides with a sufficiently low transition temperature $T_c$ [Nature **536** (2016) 309-311]. They further point out that this experimental finding is incompatible with the standard Bardeen-Cooper-Schrieffer (BCS) description. Here we report that if $T_c$ is sufficiently low, applying the renormalization group approach into the BCS action at zero temperature will naturally lead to the parabolic scaling. Our result indicates that when $T_c$ sufficiently approaches zero, quantum fluctuations will be overwhelmingly amplified so that the mean-field approximation may break down at zero temperature.

## 1. Introduction

Since the discovery of superconductivity at elevated temperature in copper-oxide materials, there has been a lot of effort to seek the possible correlations between the physical quantities controlling superconducting mechanism. One of the earliest patterns is referred to as the Uemura relation [1-2]: $T_c \propto n_s(0)/m$, which marks the linear scaling between the superfluid density at zero temperature $n_s(0)$ and the transition temperature $T_c$, where $m$ denotes the electron effective mass. The Uemura relation works reasonably well for the underdoped materials; however, it does not describe optimally doped or overdoped materials [3]. Later, Homes *et al.* found a universal scaling [3]: $T_c \propto n_s(0)/(m \cdot \sigma_{dc})$, where $\sigma_{dc}$ denotes the d.c. conductivity measured at approximately $T_c$. This scaling, which is known as "Homes' law", holds regardless of underdoped, optimally doped and overdoped materials. Thus, the

---

[†] Corresponding author.
E-mail address: taoyingyong@yahoo.com

Uemura relation can be regarded as a special case of Homes' law. It therefore attracts much attention [4-8]. Now we have been aware that Homes' law can be derived by using the Bardeen-Cooper-Schrieffer (BCS) theory [4, 8-9]. Homes *et al.* first point out [4] that Homes' law is a hallmark of dirty-limit BCS superconductor. Furthermore, Kogan argues [8] on the basis of the BCS theory that Homes' law holds not only in the extreme dirty limit but in a broad range of scattering parameter, and hence can be considered as yet another confirmation of the BCS theory.

Despite these successes, some scholar questioned the validity of Homes' law in strongly underdoped and overdoped materials. Tallon *et al.* first reported that Homes' law might break down in overdoped materials [6]. Later, some measurements on strongly underdoped materials show that the scaling of $T_c$ with $n_s(0)$ is actually sublinear [10-12]: $T_c \propto [n_s(0)]^\alpha$ with $\alpha \approx 0.5$. On the side of theoretical research, Tao recently showed that applying the renormalization group approach into the Landau-Ginzburg action combined with a Chern-Simons term might lead to an exact parabolic scaling [13]:

$$T_c \propto \sqrt{n_s(0)} \cdot \sigma_H, \qquad (1)$$

where $\sigma_H$ denotes the Hall conductivity at normal state.

Nevertheless, Tao did not clarify the range of applicability of the parabolic scaling (1). More recently, by investigating overdoped copper oxides (thin films), Božović *et al.* surprisingly observed the parabolic scaling. Concretely, they found that [14] the scaling relation between the superfluid phase stiffness at zero temperature $\rho_{so}$ and the transition temperature $T_c$ exhibits a two-class structure (please see Figure 2d in reference [14]):

$$\begin{cases} T_c \propto \rho_{so}, & T_c > T^* \\ T_c \propto \sqrt{\rho_{so}}, & T_c < T^*, \end{cases} \qquad (2)$$

where $T^*$ denotes a sufficiently low temperature.

Since the superfluid phase stiffness at zero temperature is defined by $\rho_{so} = (\hbar^2 \cdot n_s(0))/(4k_B \cdot m)$ [14], Božović *et al.* use $\rho_{so}$ to specify the superfluid density at zero temperature. Here $k_B$ is the Boltzmann constant and $\hbar$ is the reduced Planck constant. The two-class scaling (2) clearly indicates that the Uemura linear scaling holds at $T_c > T^*$, but that a parabolic scaling will emerge at $T_c < T^*$. By further measurement, Božović *et al.* investigate that the parabolic scaling is also inconsistent with Homes' law [14]. Therefore, they point out that their experimental finding is incompatible with the standard Bardeen-Cooper-Schrieffer (BCS) description. The main purpose of this paper is to show that if $T_c$ is sufficiently low, applying the renormalization group approach into the BCS action at zero temperature will naturally lead to the parabolic scaling $T_c \propto \sqrt{\rho_{so}}$. In this paper, we order $\hbar = c = k_B = 1$ with $c$ being the light speed.

Before beginning our computation, here we outline the basic idea of our method. First, we expand the order parameter $\phi$ around $T_c$ to obtain the Landau-Ginzburg

action:
$$\mathcal{L}_{LG}(T) = \mathcal{L}_0 + \lambda_2(T) \cdot |\phi|^2 + \lambda_4(T) \cdot |\phi|^4,$$
where $|T - T_c| \approx 0$. Here $\lambda_2(T)$ and $\lambda_4(T)$ are undetermined parameters. Second, by using the BCS theory, we determine the coefficients $\lambda_2(T)$ and $\lambda_4(T)$. Third, $T_c$ is assumed to be sufficiently low (i.e., $T_c \approx 0$) so that $\mathcal{L}_{LG}(T)$ is valid at $T = 0$. Thus, we apply the renormalization group procedure into $\mathcal{L}_{LG}(T = 0)$ to obtain the parabolic scaling $T_c \propto \sqrt{\rho_{so}}$. Finally, we present an explanation to illustrate why the two-class scaling (2) emerges.

## 2. BCS action at zero temperature

The starting point of this paper is the following Landau-Ginzburg action:
$$\mathcal{L}(T) = |\nabla\phi|^2 + \lambda_2(T) \cdot |\phi|^2 + \lambda_4(T) \cdot |\phi|^4, \tag{3}$$
where $|T - T_c| \approx 0$ and $\phi(x)$ is proportional to the energy gap $\Delta(x)$ [9, 15].

By using the BCS Hamiltonian of superconductivity, Gor'kov has shown that the Landau-Ginzburg equation can be written in the form [9, 15]:
$$\frac{1}{2m^*}\nabla^2\psi(x) - \frac{1}{\lambda} \cdot \frac{(T-T_c)}{T_c}\psi(x) - \frac{1}{\lambda \cdot n_s(0)}|\psi(x)|^2\psi(x) = 0, \tag{4}$$
where $\lambda = \frac{7\zeta(3)\cdot\varepsilon_F}{6\pi^2 T_c^2}$ ($\zeta(x)$ is Reimann's zeta function and $\varepsilon_F$ is the Fermi energy) and
$$\psi(x) = \left(\frac{7\zeta(3)n_s(0)}{8\pi^2 T_c^2}\right)^{\frac{1}{2}} \Delta(x). \tag{5}$$

By rescaling $\psi(x)$ according to $\phi(x) = \frac{1}{\sqrt{2m^*}}\psi(x)$, the equation (4) yields the following Lagrangian function:
$$\mathcal{L}(T) = |\nabla\phi|^2 + \frac{2m^*}{\lambda} \cdot \frac{(T-T_c)}{T_c} \cdot |\phi|^2 + \frac{2m^{*2}}{\lambda \cdot n_s(0)} \cdot |\phi|^4, \tag{6}$$

Comparing the Lagrangian functions (3) and (6) we obtain
$$\lambda_2(T) = \frac{2m^* \cdot (T-T_c)}{\lambda T_c}, \tag{7}$$

$$\lambda_4(T) = \frac{2m^{*2}}{\lambda \cdot n_s(0)}. \tag{8}$$

Here $m^*$ denotes the effective mass of a Cooper pair. Because $\lambda_2(T)$ and $\lambda_4(T)$ contain $m^*$, equations (7) and (8) imply that the contribution from quantum fluctuation and thermal fluctuation can be absorbed into $m^*$. Therefore, $m^*$ may be a function of temperature $T$; that is, $m^* = m^*(T)$. In the vicinity of $T_c$, Tao has shown [16] $m^*(T) \propto (T_c - T)^\beta$ with $\beta$ being a critical exponent. Moreover, $n_s(0)$ denotes the superfluid density at zero temperature, which should be equal to the total number density of electrons at the normal state [9].

Before beginning the renormalization group procedure, we need to clarify the range of applicability of Lagrangian (3) (or (6)). In reference [9, 15], Gor'kov derives the coefficients $\lambda_2(T)$ and $\lambda_4(T)$ by assuming a constant density of states. This simplification does not constitute a serious drawback in the mean-field theory; however, it may lead to inconsistent results if one wants to relate the bare coefficients to the effects of quantum fluctuations at criticality, where the number of spatial dimensions, $D$, may affect the critical behavior of the coefficients $\lambda_2(T)$ and $\lambda_4(T)$. In fact, a constant density of states would occur only at $D = 2$. Remarkably, Moca [17] and other scholars [18-19] have computed $\lambda_2(T)$ and $\lambda_4(T)$ at $D = 2$. Their computing results agree with Gorkov's coefficients $\lambda_2(T)$ and $\lambda_4(T)$ besides some dimensionless constants. Therefore, we conclude that Lagrangian (3) (or (6)) holds at $D = 2$. Here we also need to remind that if one expands $\mathcal{L}(T)$ in powers of the energy gap $\Delta(x)$, then the coefficient of $|\Delta(x)|^4$ is proportional to $\frac{7\zeta(3)}{8\pi^2 T_c^2}$ [17], which seems to contradict equation (8). However, once one observes the normalization condition (5), which is obtained by using the Landau-Ginzburg supercurrent equation [15], one can understand that the coefficient of $|\phi(x)|^4$ is indeed denoted by equation (8). In this paper, we mainly investigate a 2-dimensional superconducting system.

The Lagrangian (3) specified by the coefficients (7) and (8) can be regarded as the BCS action in the vicinity of $T_c$. The focus of this paper is on the BCS action at absolute zero temperature $T = 0$; that is,
$$\mathcal{L}(T = 0) = |\nabla\phi|^2 + \lambda_2(0) \cdot |\phi|^2 + \lambda_4(0) \cdot |\phi|^4. \tag{9}$$
To guarantee that $\mathcal{L}(T)$ holds at $T = 0$, we first assume $T_c \approx 0$. Since $\mathcal{L}(T = 0)$ stays at the vicinity of critical point $T_c$, the fluctuation should be so strong that applying the mean-field approach into $\mathcal{L}(T = 0)$ may be invalid. Thus, we must apply the renormalization group approach to deal with the critical behavior of $\mathcal{L}(T = 0)$. To this end, let us write down the partition function in $D$-dimensional space:
$$Z_0(T) = \int \mathcal{D}\phi^*(x) \int \mathcal{D}\phi(x) \, e^{-S_0(T)}, \tag{10}$$
where $S_0(T) = \frac{1}{T} \int d^D x \cdot \mathcal{L}(T)$ and $D = 2$.

The partition function (10) can be used to deal with the thermal critical phenomena at $T > 0$ [20]. Unfortunately, it breaks down at $T = 0$. To describe the critical behavior at $T = 0$, we need to introduce the imaginary time $\tau \in \left[0, \frac{1}{T}\right]$. Then, the order parameter $\phi$ is a function of space and imaginary time [21]; that is, $\phi = \phi(x, \tau)$. Thus, the partition function (10) can be rewritten in the form:
$$Z(T) = \int \mathcal{D}\phi^*(x, \tau) \int \mathcal{D}\phi(x, \tau) \, e^{-S(T)}, \tag{11}$$
where $S(T) = \int_0^{\frac{1}{T}} d\tau \int d^D x \cdot [\xi \cdot |\partial_\tau \phi(x, \tau)|^2 + \mathcal{L}(T)]$.

Here $\xi$ denotes a parameter determining the quantum dynamics (please see page 168 in reference [22]). Different from the case $T > 0$, thermal fluctuation will vanish at $T = 0$; therefore, the renormalization contribution to $\mathcal{L}(T = 0)$ should mainly come from quantum fluctuations. As is well known, the transition at $T = 0$ is referred to as the quantum phase transition [21]. The principal difference between quantum ($T = 0$) and thermal ($T > 0$) phase transitions is that, while dynamics is irrelevant for the latter (i.e., $\phi = \phi(x)$), it is crucial for the former (i.e., $\phi = \phi(x, \tau)$) (please see page 166 in reference [22]). It is easy to check that when the order parameter $\phi$ is independent of the imaginary time $\tau$, the quantum partition function (11) returns to the thermal partition function (10). Next we apply the renormalization group approach into the partition function $Z(T = 0)$.

## 3. Renormalization group equation

Adopting the procedure of renormalization group we first write down the quantum partition function (11) at $T = 0$:

$$Z(T = 0) = \int [\mathcal{D}\phi^*(x,\tau)]_\Lambda \int [\mathcal{D}\phi(x,\tau)]_\Lambda \, e^{-S(0)}, \tag{12}$$

where $\Lambda$ denotes the ultraviolet cut-off. The cut-off $\Lambda$ may be associated with some fundamental graininess of space-time (please see page 402 in reference [23] or see reference [24]). However, here we only regard $\Lambda$ as a physical parameter describing the quantum effect: The quantum fluctuations with length scales less than $\frac{1}{\Lambda}$ can be omitted.

Let us define

$$\phi = \phi_s + \phi_w, \tag{13}$$

where "$s$" stands for the "smooth" part and "$w$" stands for the "wriggly" part. Since $T = 0$, we assume that the "wriggly" part $\phi_w$ is mainly due to quantum fluctuations (rather than to thermal fluctuations).

Substituting equations (9) and (13) into equation (12) yields:

$$Z = \int \mathcal{D}\phi_s \int \mathcal{D}\phi_s^* \int \mathcal{D}\phi_w \int \mathcal{D}\phi_w^* \, e^{-S(0)} \tag{14}$$

If we integrate out $\phi_w$, then equation (14) can be rewritten in the form:

$$Z = \int \mathcal{D}\phi_s \int \mathcal{D}\phi_s^* \, e^{-S_{eff}(0)}, \tag{15}$$

where

$$S_{eff}(0) = \int d\tau \int d^D x \, \{(\xi + \Delta\xi) \cdot |\partial_\tau \phi_s|^2 + (1 + \Delta E) \cdot |\nabla \phi_s|^2 + (\lambda_2(0) + \Delta\lambda_2) \cdot |\phi_s|^2 + (\lambda_4(0) + \Delta\lambda_4) \cdot |\phi_s|^4\}. \tag{16}$$

Here $\Delta\lambda_2$, $\Delta\lambda_4$, $\Delta\xi$ and $\Delta E$ stand for perturbative terms (please refer to page 400 in [23]), which absorb the contributions from quantum fluctuations. In terms of the rescaled variable $x' = b^{-1}x$ and $\tau' = b^{-z}\tau$, equation (16) becomes

$$S_{eff}(0) = \int d\tau' \int d^D x' \, \{(\xi + \Delta\xi) \cdot b^{D-z} \cdot |\partial'_\tau \phi_s|^2 + (1 + \Delta E) \cdot b^{D+z-2} \cdot |\nabla' \phi_s|^2 + (\lambda_2(0) + \Delta\lambda_2) \cdot b^{D+z} \cdot |\phi_s|^2 + (\lambda_4(0) + \Delta\lambda_4) \cdot b^{D+z} \cdot |\phi_s|^4\}, \tag{17}$$

where $z$ denotes the quantum dynamical exponent [21].

If we rescale the field $\phi_s$ according to $\phi'_s = [b^{D+z-2} \cdot (1 + \Delta E)]^{\frac{1}{2}} \cdot \phi_s$, then equation (17) can be rewritten in the form:

$$S_{eff}(0) = \int d\tau' \int d^D x' \, \{\xi' \cdot |\partial'_\tau \phi'_s|^2 + |\nabla' \phi'_s|^2 + \lambda'_2(0) \cdot |\phi'_s|^2 + \lambda'_4(0) \cdot |\phi'_s|^4\},$$

(18)

which shares the same form with the Lagrangian (9).

The new parameters of the action (18) are:
$$\xi' = (\xi + \Delta\xi)(1 + \Delta E)^{-1} b^{2-2z}, \tag{19}$$
$$\lambda_2'(0) = (\lambda_2(0) + \Delta\lambda_2)(1 + \Delta E)^{-1} b^2, \tag{20}$$
$$\lambda_4'(0) = (\lambda_4(0) + \Delta\lambda_4)(1 + \Delta E)^{-2} b^{4-D-z}. \tag{21}$$

All of the corrections, $\Delta\lambda_2$, $\Delta\lambda_4$, $\Delta\xi$ and $\Delta E$, should be small compared to the leading terms if perturbation theory is justified. It is easy to check that the Lagrangian (9) is a $O(2)$-symmetric $\phi^4$ theory. Therefore, if one keeps the perturbative terms in equations (20) and (21) up to the one-loop correction, one obtains (please see page 169 in reference [22]):

$$\lambda_2'(0) = b^2 \left[ \lambda_2(0) + 4\lambda_4(0) \int_{-\infty}^{\infty} \frac{d\omega}{2\pi} \int_{\Lambda/b}^{\Lambda} \frac{d^D k}{(2\pi)^D} \left( \frac{1}{\omega^2 + k^2 + \lambda_2(0)} - \frac{1}{\omega^2 + k^2} \right) \right], \tag{22}$$

$$\lambda_4'(0) = b^{4-D-z} \lambda_4(0) \left[ 1 - 10\lambda_4(0) \int_{-\infty}^{\infty} \frac{d\omega}{2\pi} \int_{\Lambda/b}^{\Lambda} \frac{d^D k}{(2\pi)^D} \left( \frac{1}{\omega^2 + k^2 + \lambda_2(0)} \right)^2 \right], \tag{23}$$

where $\omega$ and $k$ denote Matsubara frequency and momenta, respectively.

If we define the dimensionless coupling as

$$\hat{\lambda}_4(0) = \lambda_4(0) \cdot \Lambda^{D+z-4} \cdot \frac{(\pi)^{\frac{D}{2}}}{2(2\pi)^D \Gamma\left(\frac{D}{2}\right)}, \tag{24}$$

equations (22) and (23) yield the following renormalization group equations:

$$\frac{d\lambda_2(0)}{d\ln b} = \lambda_2(0) \cdot \left(2 - 4\hat{\lambda}_4(0)\right) + O(\hat{\lambda}_4^2). \tag{25}$$

$$\frac{d\hat{\lambda}_4(0)}{d\ln b} = (4 - D - z) \cdot \hat{\lambda}_4(0) + 10\hat{\lambda}_4^2(0) + O(\hat{\lambda}_4^3). \tag{26}$$

## 4. Results

By using equations (25) and (26), it is easy to obtain a nontrivial fixed point of the renormalization group flows as:

$$\hat{\lambda}_4^*(0) \approx \frac{4-D-z}{10}. \tag{27}$$

The fixed point $\hat{\lambda}_4^*(0)$, which is different from the Wilson-Fisher fixed point [25], holds when $D < 4 - z$. Therefore, the upper critical dimension is equal to $4 - z$. The renormalization group equations (25) and (26) indicate that the Lagrangian (9) will vary as the length scaling changes. The change in the Lagrangian (9) comes from the contributions of quantum fluctuations at different scales. However, if the critical point $T_c$ tends to 0, the Lagrangian (9) will evolve to a stable form denoted by the fixed point $\hat{\lambda}_4^*(0)$, which describes the critical behavior around $T_c$.

Substituting equation (27) into equation (24) we obtain:

$$\lambda_4^*(0) \approx (4 - D - z) \cdot \Lambda^{4-D-z} \cdot \frac{(2\pi)^D \Gamma\left(\frac{D}{2}\right)}{5(\pi)^{\frac{D}{2}}}. \tag{28}$$

To investigate the relation between the superfluid phase stiffness at zero temperature $\rho_{so}$ and the transition temperature $T_c$, let us define [14]:

$$\rho_{so} = \frac{n_s(0)}{2m^*(T=0)}. \tag{29}$$

Substituting equation (29) into equation (8) at $T = 0$ yields:

$$\lambda_4(0) = \frac{6\pi^2 m^*(0) \cdot T_c^2}{7\zeta(3) \cdot \varepsilon_F \cdot \rho_{so}}. \tag{30}$$

Near the critical point $T_c$, $\lambda_4(0)$ becomes $\lambda_4^*(0)$; therefore, substituting equation (30) into equation (28) we get:

$$T_c = \sqrt{(4-D-z) \cdot \Lambda^{4-D-z} \cdot \frac{7(2\pi)^D \Gamma\left(\frac{D}{2}\right) \zeta(3) \cdot \varepsilon_F}{30(\pi)^{\frac{D}{2}+2} m^*(0)}} \cdot \sqrt{\rho_{so}}. \tag{31}$$

By dimensional analysis one has $[n_s(0)] = [\Lambda]^D$ and $[m^*(0)] = [\varepsilon_F] = [T_c] = [\Lambda]$, so we have $[\rho_{so}] = [\Lambda]^{D-1}$. Thus, equation (31) implies the following equation:

$$[\Lambda] = [\Lambda]^{\frac{4-D-z}{2}} \cdot [\Lambda]^{\frac{D-1}{2}}. \tag{32}$$

By equation (32) we obtain $z = 1$. Therefore, equation (31) can be rewritten in the form:

$$T_c = \sqrt{(3-D) \cdot \Lambda^{3-D} \cdot \frac{7(2\pi)^D \Gamma\left(\frac{D}{2}\right) \zeta(3) \cdot \varepsilon_F}{30(\pi)^{\frac{D}{2}+2} m^*(0)}} \cdot \sqrt{\rho_{so}}. \tag{33}$$

Different from $m^*(T > 0)$, $m^*(0)$ does not receive the contribution from thermal fluctuations, because all the thermal effect vanish at the absolute zero temperature $T = 0$. In contrast, quantum fluctuations will emerge at the absolute zero temperature. Due to the absence of thermal fluctuations, $m^*(0)$ mainly receives the contributions from quantum fluctuations at $T = 0$. Because there is no thermal effect at $T = 0$, the value of $m^*(0)$ will be fixed at the fixed point $\hat{\lambda}_4^*(0)$. Thus, the scaling law (33) reproduces the experimental result $T_c \propto \sqrt{\rho_{so}}$ for $D = 2$ [14]. The scaling law (33) is the main result of this paper. However, it may break down once $T_c$ is sufficiently larger than 0. There are two reasons. First, if $T_c$ is sufficiently larger than 0, by our preceding discussion, the Lagrangian (9) is invalid (because the Lagrangian (9) holds if and only if $T_c \approx 0$). Thus, our scaling law (33) has no meaning. Second, if $T_c$ is sufficiently larger than 0, by equation (3) we will have $T \approx T_c \gg 0$. Then, the effect of thermal fluctuation cannot be omitted; therefore, $m^*(T)$ will receive the contribution from thermal fluctuation. Thus, $m^*(T)$ have to depend on the temperature $T$ so that our scaling law (33) is destroyed. If we denote $T^*$ by the characteristic temperature below which the quantum effect dominates the fluctuation, then we understand that our scaling law (33) holds at $T_c < T^*$. This is in accordance with the parabolic scaling in the experimental result (2).

It is worth mentioning that we cannot eliminate the possibility that equation (33) approximately holds for $D = 3$, since the constant density of states is a good approximation around the Fermi surface. If equation (33) indeed holds for $D = 3$, then it predicts a strange result: $T_c = 0$. To understand this result, let us return to equation (27) which indicates that $D = 3$ is the upper critical dimension. This means that the mean-field approximation is valid at $D \geq 3$, where quantum fluctuations can be averaged out. From this meaning, equation (33) actually predicts that the BCS quantum critical behavior for $D = 3$ only occurs at $T_c = 0$, and hence it is trivial. In contrast, a non-trivial BCS quantum critical behavior should occur at $D = 2$. It seems to resemble the Kosterlitz-Thouless transition [26], which also occurs at $D = 2$. Because the mean-field approximation is valid at $D \geq 3$, let us start to derive the relationship between $T_c$ and $\rho_{so}$ by using the mean-field approximation.

## 5. Revisiting Homes' law

To explain the linear scaling in the experimental result (2), we next show that Homes' law can be derived by using the mean-field approach.

In isotropic BCS superconductors, by using the mean-field approximation the penetration depth is given by [8-9]

$$\lambda_p(T) = \frac{1}{\sqrt{4\pi Q(T)}}, \tag{34}$$

where

$$Q(T) = \frac{Ne^2}{m} 2\pi T \Delta^2 \sum_{n=1}^{\infty} \frac{1}{[(2n+1)^2\pi^2 T^2 + \Delta^2]\left(\sqrt{(2n+1)^2\pi^2 T^2 + \Delta^2} + \frac{1}{2\tau_s}\right)}.$$

Here $\tau_s$ denotes the scattering relaxation time, $e$ denotes the electron charge and $N$ denotes the total number density of electrons at the normal state.

If one orders $u = (2n+1)\pi T$, one has:

$$\lim_{T \to 0} 2\pi T \sum_{n=1}^{\infty} \Rightarrow \int_0^{\infty} du. \tag{35}$$

On the other hand, equation (34) can be rewritten in the form:

$$\lambda_p^{-2}(T) = 4\pi \frac{Ne^2}{m} 2\pi T \Delta^2 \sum_{n=1}^{\infty} \frac{1}{[(2n+1)^2\pi^2 T^2 + \Delta^2]\left(\sqrt{(2n+1)^2\pi^2 T^2 + \Delta^2} + \frac{1}{2\tau_s}\right)}. \tag{36}$$

Here we assume that the following limit exists:

$$\lim_{T \to 0} \lambda_p^{-2}(T) = \lambda_p^{-2}(0). \tag{37}$$

Thus, $\lambda_p(0)$ denotes the penetration depth at zero temperature.

Using equations (35) and (37), imposing $T \to 0$ on equation (36) yields:

$$\lambda_p^{-2}(0) = 4\pi \frac{Ne^2}{m} \Delta(0)^2 \int_0^{\infty} \frac{1}{(u^2 + \Delta(0)^2)\left(\sqrt{u^2 + \Delta(0)^2} + \frac{1}{2\tau_s}\right)} du, \tag{38}$$

where $\Delta(0) \propto T_c$ denotes the energy gap at zero temperature.

Computing the integral (38) one obtains:

$$\lambda_p^{-2}(0) = 4\pi^2 \sigma_{dc} \Delta(0) \cdot \left(1 - \frac{4}{\pi\sqrt{1-\epsilon^2}} tan^{-1}\sqrt{\frac{1-\epsilon}{1+\epsilon}}\right), \tag{39}$$

where $\epsilon = \frac{1}{2\tau_s \Delta(0)}$, $\sigma_{dc} = \frac{N\tau_s e^2}{m}$ and we have used the formula [27]:

$$\int_0^{\infty} \frac{dx}{(1+x^2)(\sqrt{1+x^2}+y)} = \frac{\pi}{2y} - \frac{2}{y(1-y^2)} tan^{-1}\sqrt{\frac{1-y}{1+y}}. \tag{40}$$

Since $\rho_{so} \propto \lambda_p^{-2}(0)$ and $\Delta(0) \propto T_c$, from equation (40) we obtain Homes' law:

$$T_c \propto \frac{\rho_{so}}{\sigma_{dc}}, \tag{41}$$

which was also obtained by Kogan [8].

The above approach to derive the scaling (41) was first performed by Abrikosov, Gor'kov and Dzyaloshinskii [9], and was recently developed by Kogan [8]. Here the purpose of combining their computation processes is to check the range of applicability of the scaling (41). By the discussion above, the key step of obtaining the scaling (41) is to impose the limit $T \to 0$ on equation (36). The validity of this step is based on the mean-field approximation: All the quantum fluctuations and thermal fluctuations can be averaged out, so the infinite summation operation in equation (36) is enough fine at $T \to 0$; thus, we can transform the infinite summation into an integral, i.e., formula (38) holds. However, when $D = 2$, these deductions hold only at $T_c > T^*$. To see this, let us remind that by the definition of $T^*$ we should have $T^* \to 0$. Since near a critical point the fluctuation will be overwhelmingly amplified, $T_c < T^*$ implies that quantum fluctuations at $T = 0$ will become extremely strong, so that the mean-field approximation may break down at $T = 0$. Therefore, we conclude that if $D = 2$, the scaling (41) holds at $T_c > T^*$.

The analyses above are based on $D = 2$. Then, the mean-field result (41) remarkably differs from the renormalization-group result (33). However, since $D = 3$ denotes the upper critical dimension, the mean-field result (41) should be always a good description for $T_c \geq 0$ when $D = 3$. This implies an experimentally testable prediction that, for 3-dimensional BCS superconducting materials, the two-class scaling (2) will be replaced by the linear scaling (41).

## 6. Discussion and Conclusion

We have shown that the linear scaling $T_c \propto \rho_{so}$ in the experimental result (2) (at $T_c > T^*$) can be regarded as a mean-field result. This means that when the transition temperature $T_c$ is far away from 0, the mean-field approximation may be valid at zero temperature. Conversely, if the transition temperature $T_c$ approaches 0, by using the renormalization group approach we show that a parabolic scaling $T_c \propto \sqrt{\rho_{so}}$ will emerge. Since the thermal effect vanishes at zero temperature, the renormalization effect should be mainly due to quantum fluctuations. Therefore, our work implies that one can artificially adjust the transition temperature $T_c$ (to approach zero) to stimulate quantum fluctuations, just as Božović *et al.* have experienced [14]. It is worth mentioning that our result differs from the current quantum critical result [7, 28], which predicts the linear scaling $T_c \propto \rho_{so}$ for the thin superconducting film ($D = 2$). Unfortunately, the experimental result (2) indicates that, when $T_c < T^*$, $T_c \propto \sqrt{\rho_{so}}$ rather than $T_c \propto \rho_{so}$. Because of this fact, Božović *et al.* further conclude that their experimental findings challenge the existing theories [14]. However, by applying the renormalization group approach into the BCS action at zero temperature we reproduce their experimental result at $T_c < T^*$. In fact, Božović *et al.* obtain the experimental result (2) by testing the overdoped copper oxides, which are perceived as simpler, with strongly correlated fermion physics evolving smoothly into the conventional BCS behavior [14]. This is a possible reason why applying the renormalization group approach into the BCS theory can produce their experimental result. In contrast, we question the validity of the scaling law (33) in underdoped materials, where BCS theory may break down [29].

In conclusion, we have produced the two-class scaling (2) within the framework of BCS theory. When the transition temperature $T_c$ is enough larger than zero, the quantum fluctuation at zero temperature can be approximately averaged out. Thus, the mean-field approach successfully produces the linear scaling in the experimental result (2) with $T_c$ being relatively high. However, once $T_c$ approaches zero, the quantum fluctuation at zero temperature will be overwhelmingly amplified so that the mean-field approximation is no longer valid. Remarkably, applying the renormalization group approach into the BCS theory at zero temperature produces the parabolic scaling in the experimental result (2) with $T_c$ being sufficiently low. This means that when the transition temperature $T_c$ approaches absolute zero, a quantum critical phenomenon (e.g., parabolic scaling) might emerge. Since our theory is based on the BCS framework, the parabolic scaling can be regarded as a BCS quantum critical phenomenon. As a potential application, the parabolic scaling may be also thought of as a signal capturing quantum fluctuations.

## Acknowledgments

This work was supported by the Fundamental Research Funds for the Central Universities (Grant No. SWU1409444).